\documentclass[apj]{emulateapj}
\slugcomment{{\sc ApJ Accepted:} July , 6 2011}
\pdfoutput=1
\usepackage{graphicx,natbib,avant,color,psfrag,amsmath}
\usepackage{multirow, array,booktabs,mdwlist}

\shorttitle{Spectral characterization of FU Orionis S}

\begin{document}

\title{Constraining mass ratio and extinction in the FU Orionis binary system with infrared integral field spectroscopy}



\author{Laurent Pueyo \altaffilmark{1}, Lynne Hillenbrand \altaffilmark{2}, Gautam Vasisht \altaffilmark{3}, Ben R. Oppenheimer \altaffilmark{4}, John D. Monnier \altaffilmark{5}, Sasha Hinkley \altaffilmark{2}, Justin Crepp \altaffilmark{7},  Lewis C. Roberts Jr\altaffilmark{3}, Douglas Brenner \altaffilmark{4}, Neil Zimmerman \altaffilmark{4},  Ian Parry \altaffilmark{8}, Charles Beichman \altaffilmark{6}, Richard Dekany \altaffilmark{2}, Mike Shao \altaffilmark{3}, Rick Burruss \altaffilmark{3}, Eric Cady \altaffilmark{3}, Jenny Roberts \altaffilmark{2}, R\'{e}mi Soummer \altaffilmark{1}}

\altaffiltext{1}{Space Telescope Science Institute, 3700 San Martin Drive, Baltimore, MD 21218}
   
\altaffiltext{2}{Department of Astronomy, California Institute of Technology, 1200 E. California Blvd., Pasadena, CA 91125} 

\altaffiltext{3}{Jet propulsion Laboratory, California Institute of technology, 4800 Oak Grove Drive, Pasadena, CA 91109 , USA} 

\altaffiltext{4}{American Museum of Natural History, Central Park West at 79th Street, New York, NY 10024}

\altaffiltext{5}{Department of Astronomy, University of Michigan, 941 Dennison Building, 500 Church Street, Ann Arbor, MI 48109-1090}
  
\altaffiltext{6} {NASA Exoplanet Science Institute, 770 S. Wilson Avenue, Pasadena, CA 91225}
 
 \altaffiltext{7} {Department of Physics, 225 Nieuwland Science Hall, University of Notre Dame, Notre Dame, IN 46556, USA }
 
 \altaffiltext{8}{University of Cambridge, Institute of Astronomy, Madingley Rd, Cambridge, CB3 0HA, UK}

%
%


\begin{abstract}
We report low resolution near infrared spectroscopic observations of the eruptive star FU Orionis using the Integral Field Spectrograph Project 1640 installed at the Palomar Hale telescope. This work focuses on elucidating the nature of the faint source, located $0.5''$ south of FU Ori, and identified in 2003 as FU Ori S. We first use our observations in conjunction with published data to demonstrate that the two stars are indeed physically associated and form a true binary pair. We then proceed to extract J and H band spectro-photometry using the damped LOCI algorithm, a reduction method tailored for high contrast science with IFS. This is the first communication reporting the high accuracy of this technique, pioneered by the Project 1640 team, on a faint astronomical source. We use our low resolution near infrared spectrum in conjunction with $10.2$ micron interferometric data to constrain the infrared excess of FU Ori S. We then focus on estimating the bulk physical properties of FU Ori S. Our models lead to estimates of an object heavily reddened, $A_V =8-12$, with an effective temperature of $\sim$ 4000-6500 K . Finally we put these results in the context of the FU Ori N-S system and argue that our analysis provides evidence that FU Ori S might be the more massive component of this binary system.
\end{abstract}

\maketitle 

\section{Introduction}
FU Orionis is the prototype of a small class of rare, eruptive young stars. ``FUOr'' outbursts are generally interpreted as episodes of significantly enhanced accretion in the early stages of disk evolution in young low-mass T Tauri stars. In 1936, the apparent magnitude of FU Orionis brightened by 6 magnitudes and the object has been slowly fading since then \citep{1966VA......8..109H,1977ApJ...217..693H,1985PAZh...11..846K,1988ApJ...325..231K,2005MNRAS.361..942C}. Only a few other sources have been identified to undergo similar eruptive phenomena with V1057 Cyg and V1515 Cyg \citep{1991ApJ...383..664K,1975PASP...87..379L,1977PASP...89..704L} serving along with FU Orionis as the classical examples. 

In the optical, the absorption spectra of FUOr objects resemble broadened F-G type supergiants along with strong wind/outflow signatures in Balmer and other lines (e.g. CaII, Na I, K I, Li I); and in the near-infrared they have spectral absorption features similar to broadened K-M supergiant atmospheres. They generally lack the rich emission-line spectrum that is characteristic of rapidly accreting young low mass stars.  FUOr objects also exhibit strong infrared through millimeter excesses, consistent with the presence of an accretion disk and/or a circumstellar envelope of dust.  Based on these diagnostics, about a dozen additional sources have been identified as ``FU Ori-like'', with a growing number of them known as wide binary systems (e.g. L1551 IRS5, \citet{1998Natur.395..355R}, RNO 1B/C \cite{1993AJ....105.1505K}, AR  6A/B \cite{2003AJ....126.2936A}), including FU Ori
\citep{2004ApJ...601L..83W}. For a more exhaustive description of the observational properties of FU Ori stars, we direct the reader towards the review papers by \citet{1996ARA&A..34..207H} and \citet{2010vaoa.conf...19R}. 


The accretion disk model for FUOr systems, and FU Orionis itself in particular, has been constrained using interferometric measurements and developed to a sophisticated level by \citet{2007ApJ...669..483Z,2010ApJ...713.1134Z}. These authors consider broad-band photometry, spectrophotometry, and high dispersion spectroscopy to derive a model size for the outbursting region of the inner disk extending outward to 0.5-1.0 AU, with an accretion rate of $\dot{M} = 2\times 10^{-4}~M_{\odot} \; yr^{-1}$ onto a central star of mass $0.3 M_{\odot}$.  Interferometric measurements by \citet{2011arXiv1106.1440E}, building on those of \citet{2005A&A...437..627M} and \citet{2006ApJ...641..547M}, confirm the rotating accretion disk model for this source. Chandra characterization of FU Orionis unveiled a hot heavily absorbed variable emission component; the variability is thought to be the signature of coronal emission in the close vicinity of the star, seen through the accreting gas \citep{2010ApJ...722.1654S}. This interpretation seems consistent with the recently reported  presence of strong magnetic field in the innermost region of the accretion disk by \citet{2005Natur.438..466D}, and reinforces the explanation of the outburst which invokes a major increase in the surface brightness of an accretion disk because of sudden increase in the accretion rate \citep{1996ARA&A..34..207H}. 

An alternative to the accretion disk model for FUOrs, is the hypothesis of a fast rotating G-supergiant photosphere. \citet{2003ApJ...595..384H} also showed that the optical and infra-red spectroscopic features of fast rotating G-supergiants can reproduce FUOr observations if the ``boxy" line profile shapes can be attributed to core line emission rather than to a rotating disk. In an effort to reconcile both models \citet{2007AstL...33..755K} recently proposed a modified accretion disk model with a puffed inner edge, that is consistent with  the HST/STIS spectrum of FU Orionis. \citet{Hartmann2012} however fit the \citet{2007ApJ...669..483Z} disk model with $A_V = 1.5$ mag to the same UV spectrum though \citet{2007ApJ...669..483Z}  comment on the blue excess shortward of 2500-2600 \AA. Since the star/disk source is also embedded in a dust shell, extinction plays a crucial role in the spectral characterization of  FU Orionis at short wavelengths, and \citet{2007AstL...33..755K} suggested using the known companion to FU Orionis as a reference to obtain a model independent estimate of the optical extinction.

FU Ori S, a stellar companion located $0.4 ''$ to the south west of FU Ori, was first discovered by \citet{2004ApJ...601L..83W} using adaptive optics imaging at K-band. \citet{2004ApJ...601L..83W} discussed, based on statistical arguments using on J and K colors, that the discovered object was most likely gravitationally bound to FU Ori. Follow up AO observations by \cite{2004ApJ...608L..65R} did not establish common proper motion due to the large uncertainties on the proper motion of FU Orionis (a result of the large distance to FU Ori $\sim 450$ pc \cite{1977MNRAS.181..657M}) but provided further evidence of the young age of FU Ori S based on infrared excess inferred from JHK'L' colors.  Spectral line diagnostics in the K-band suggest a late G or K spectral type if the inference of \cite{2004ApJ...608L..65R} is correct that Na I and Ca I are present in their spectrum while CO absorption is lacking. Speckle interferometry observations at $0.8$ micron by \cite{2008AstBu..63..357K} find $A_{V} =2.2$  mag towards FU Ori S, provided the star is of spectral type G9 or later. The temperature of the best fitting (unreddened) blackbody to the K-band continuum was $\sim 2500 \; K$ and, based on this, it was suggested that FU Ori S exhibits a considerable infrared excess. L-band photometry is also available from these authors. \citet{2012AJ....143...55B} recently reported high spectral resolution J H and K band ($R \sim 3000$) spectra of FU Ori S, obtained with the Gemini NIFS integral field spectrograph. Based on line diagnostics they showed that FU Ori S was a highly accreting K5 type star and suggested that this object might be the more massive component of the system. N-band interferometric measurements by \citet{2009ApJ...700..491M} detected the companion, which confirms a substantial infrared excess, but no flux density was reported.  \cite{2010ApJ...722.1654S} identified a centroid offset towards FU Ori S in the soft X-ray component of Chandra observations, and argued that the most likely explanation was that FU Ori S was a weak soft X ray emitter.  All of the evidence is consistent with FU Ori S being a young low-mass, perhaps K-type, T-Tauri star. 

Precision astrometry on follow up AO observation of FU Ori, with a baseline longer than the time elapsed between the \citet{2004ApJ...601L..83W} and \cite{2004ApJ...608L..65R} epochs can firmly establish whether or not FU Ori and FU Ori S are gravitationally bound. Obtaining a well-constrained SED of this object in these spectral regions would allow characterization of FU Ori S independently of FU Ori N. In particular, SED measurements covering the spectral region expected to be dominated by stellar (as opposed to circumstellar) flux can provide an extinction estimate.  This measurement may or may not be applicable to both components of the binary (FU Ori N as well as to FU Ori S) depending on the geometry of the circumstellar and circumbinary material. Independent knowledge of the extinction to FU Ori N is necessary to validate the findings based on accretion disk models \citep{2007AstL...33..755K,2007ApJ...669..483Z} in which the extinction is either a derived parameter of the models or a required input to them.  Understanding the extinction therefore provides helpful insights concerning the nature of the primary (FU Ori) star as well as the geometry of its surroundings, including the companion.

In an effort to further elucidate the nature of FU Ori S by quantifying the spectral type and the reddening, and to assist in interpretation of the dust shell encircling the FU Orionis N/S system, we observed this source with the Project 1640 (P1640) Integral Field Spectrograph (IFS). We first summarize our observations, our data reduction and establish the binary nature of the FU Orionis system based on our data. Then we reconstruct the $0.8 $ to  $10 $ micron SED using previously published observations and discuss our photometric points in the context of the literature. Based on this reconstructed SED we seek to characterize the near infrared emission of FU Ori S. Our models lead to estimates of an object heavily reddened, A$_V$=8-12, with an effective temperature of $\sim$ 4000-6500 K. We furthermore quantify the amplitude of the infrared excess produced by  circumstellar dust around FU Ori S, using our J and H SED as a photometric baseline for interferometric data published by \cite{2009ApJ...700..491M}, reprocessed for the purpose of this paper. We finally put these results in perspective in the context of the FU Ori N-S system and argue that our analysis provides evidence that FU Ori S might be the hotter and therefore more massive component.

\section{Observations and data reduction}
\subsection{Observations}


FU Ori was observed at Palomar on March 17 th 2009 within the $\sim 4" \times 4"$ field of view of the P1640 Integral Field Spectrograph. The IFS design prioritizes, for a given Field Of View, fine spatial sampling over spectral resolution for  (when compared to other microlens based ISF such as OSIRIS \citet{mcelwain-2007-656}). This design provides the chromatic information necessary to discriminate faint point sources from optical artifacts \citep{2002ApJ...578..543S,JustinSpeckle} in high contrast observations.  Specifically, the IFS features a moderate spectral resolution ($R \sim 45$) and high spatial resolution ($\lambda/D=$45-72 mas). Detail of the instrument can be found in \cite{2011PASP..123...74H}. The data consist of twelve exposures of 127 seconds with the primary star occulted (behind the coronagraphic mask) and two 2  second exposures with the primary star unobscured (off the coronagraphic mask). Each exposure produces a set of $250 \times 250$ spectra on the infrared detector, each spectrum corresponding to the dispersed image of a micro-lens in the focal plane of the IFS. This raw data is then converted to a cube of $23$ image slices, each slice corresponding to the image of the focal plane microlens array at a given wavelength. The wavelength solution of the spectrograph is derived using a set of calibration frames obtained prior to and during the observing sequence \citep{NeilPipeline}.  Determination of the spectrograph response in the particular case of FU Ori is detailed in \S.~\ref{sec::SpectralCalibration}.

\begin{figure}
\includegraphics[width=8cm]{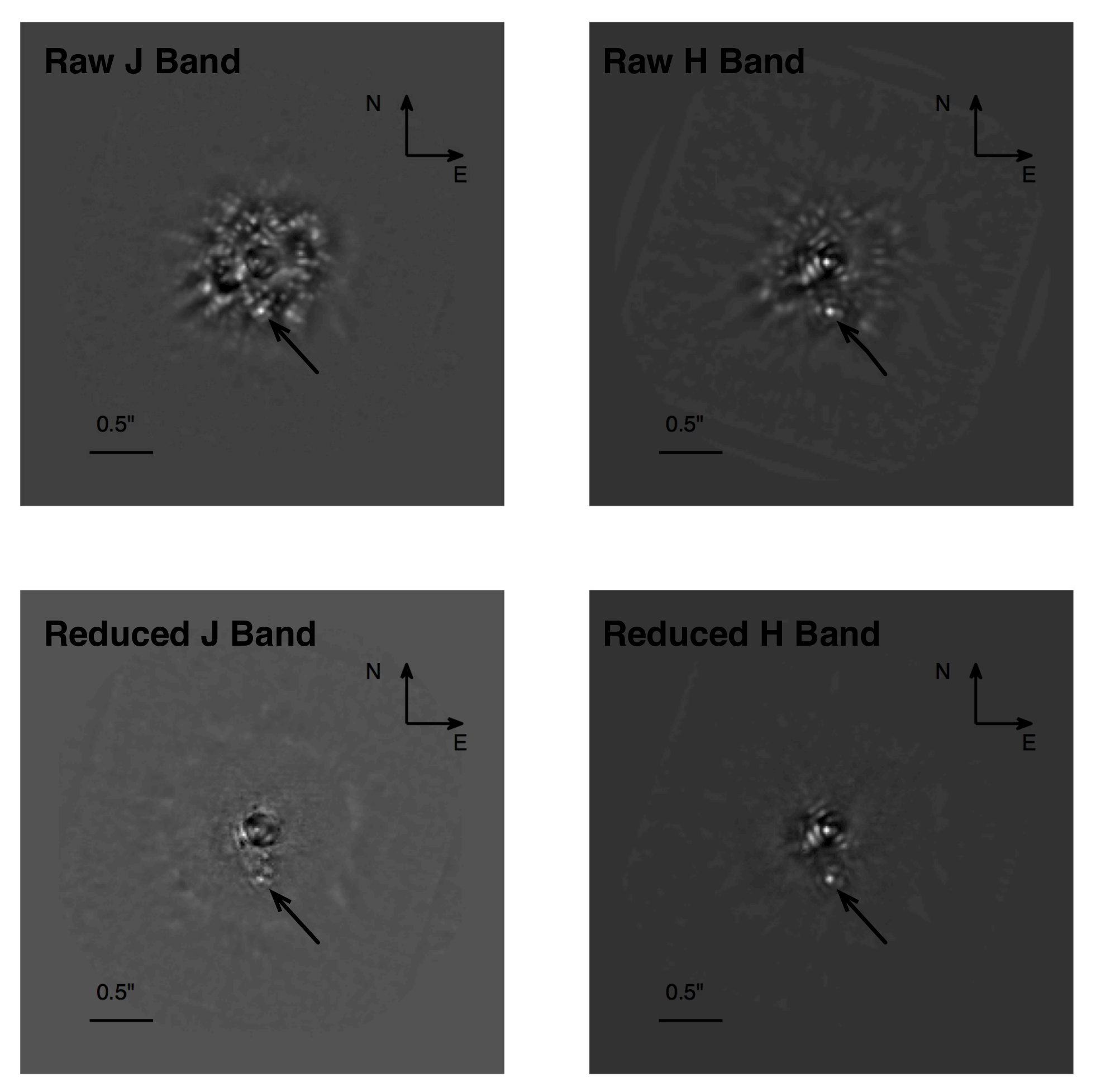}
\caption[Images]{{\em Top}: band averaged P1640 images of the FU Orionis system, FU Ori N is occulted by the coronagraphic mask. {\em Top Left}: J band. {\em Top Right}: H band. FU OriS can be identified in the raw H band images, but is as bright as the speckles in J band. Photometric estimates based on the raw IFS cube will overestimate the flux of FU Ori S because of speckle contamination. {\em Bottom}: reduced images, the detectability of FU Ori S is significantly enhanced.}
\label{fig::ImagesFuOri}
\end{figure}

The FU Ori system is seen through an optical train consisting of the Palomar Adaptive Optics system (PALAO), the Apodised Pupil Lyot Coronagraph \citep{2009ApJ...695..695S} and the P1640 spectrograph \citep{2011PASP..123...74H}. The calibrated data (data cubes with speckles that have not been suppressed by post-processing), with slices integrated over the J and the H bands are shown on the top panel of Fig.~\ref{fig::ImagesFuOri}.  We detect FU Ori S at a separation of $0.491'' \pm  0.007$ and a position angle of $ 161.2^{\circ} \pm     1.1$ with respect to FU Ori N.  The contrast sensitivity of the data are to $\Delta M_H =6.5 $ mag and  $\Delta M_J = 6.2$  mag for companions at $0.5''$ separation, somewhat worse contrast than typical of P1640 contrast curves (e.g. \citet{JustinSpeckle}) due to the early stage of the data.

While the companion FU Ori S can be identified in H band, it is as bright as the speckles in J and thus can only be distinguished from them using the chromatic diversity provided by the IFS. Recent upgrades of the Palomar high contrast near-infrared system, including a new Adaptive Optics systems \citep{dekany:62720G} and an interferometric wavefront calibration system \citep{2010lyot.confE..51V,wallace:74400S}, will further improve the speckle noise level. Direct extraction of the FU Ori S spectrum using the raw slices  that were integrated to produce the top panel of  Fig.~\ref{fig::ImagesFuOri}, leads to over-estimating the companion's spectro-photometry because of the speckles' outward motion at the location of the object as the wavelength increases. 

In order to mitigate for this effect we sought to improve the Signal to Noise Ratio (SNR) on FU Ori S using the P1640 speckle calibration pipeline presented in \cite{JustinSpeckle}. This method takes advantage of the chromatic diversity of the IFS  \citep{2002ApJ...578..543S} and combines it with optimal Point Spread Function (PSF) subtraction algorithms, e.g. LOCI; \citep{2007ApJ...660..770L}. The processed images are shown in the bottom panels of Fig.~\ref{fig::ImagesFuOri} and demonstrate that our reduction approach considerably increases the SNR of FU Ori S. 
%

\subsection{IFS spectral calibration}
\label{sec::SpectralCalibration}
Calibrating simultaneously the wavelength solution, the atmospheric dispersion and the dispersion intrinsic to the instrument with the low spectral resolution ($R \sim 45$) of P1640 is a delicate exercise. Indeed, since the telluric sky lines are averaged over each wavelength channel, they can not be used as a reference to derive the wavelength solution \citep{2004PASP..116..362C}. We thus first calibrate the wavelength solution off-line, using a laser tunable source and follow the procedure detailed in \cite{NeilPipeline}. Moreover, the relatively small field of view of P1640 prevents us from simultaneously obtaining observations of a calibration star in a science exposure and the presence of the coronagraph prevents us from using the primary star as the calibrator.

\begin{figure}
\begin{center}
\begin{tabular}{c}
\includegraphics[width=6cm]{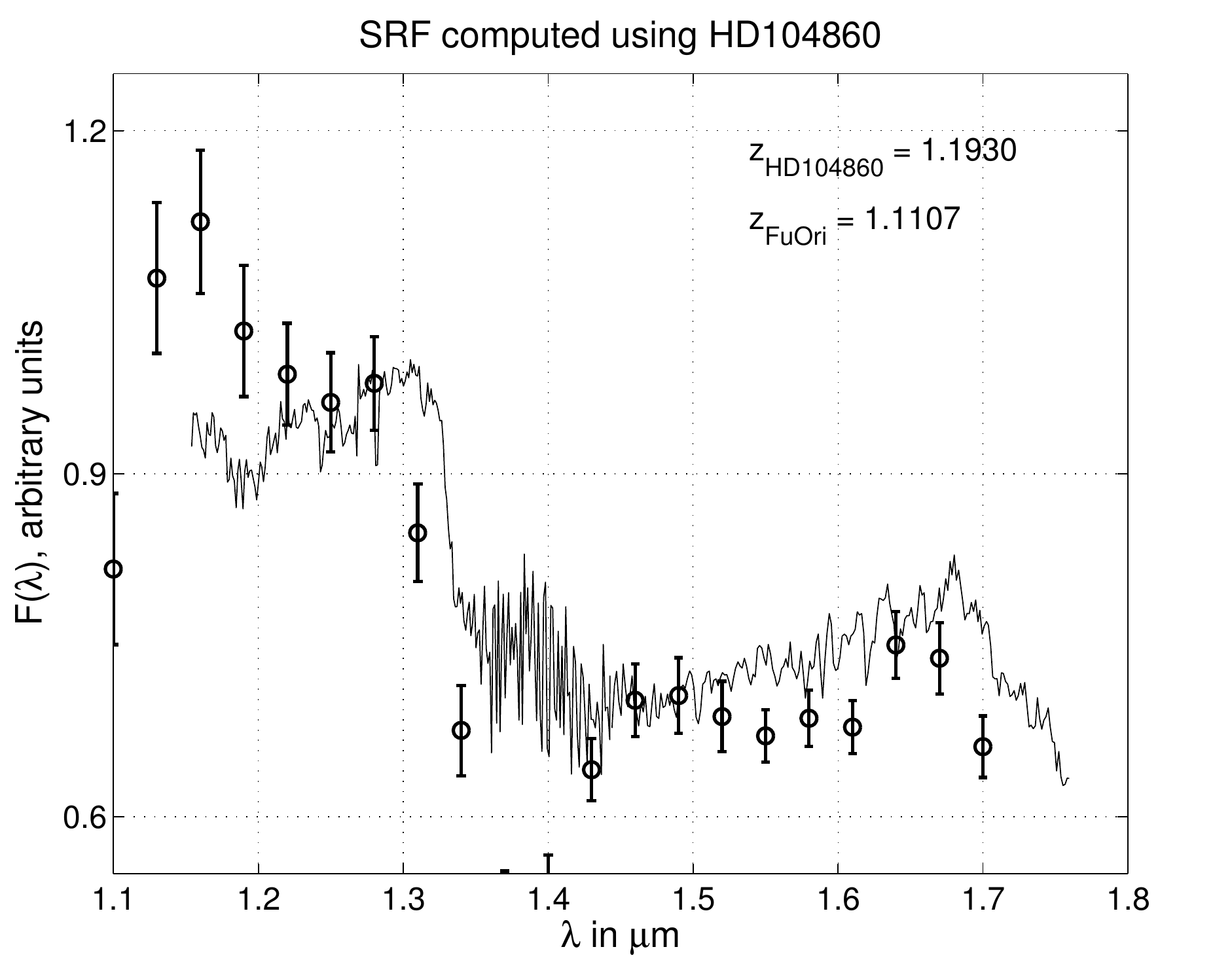}\\
\includegraphics[width=6cm]{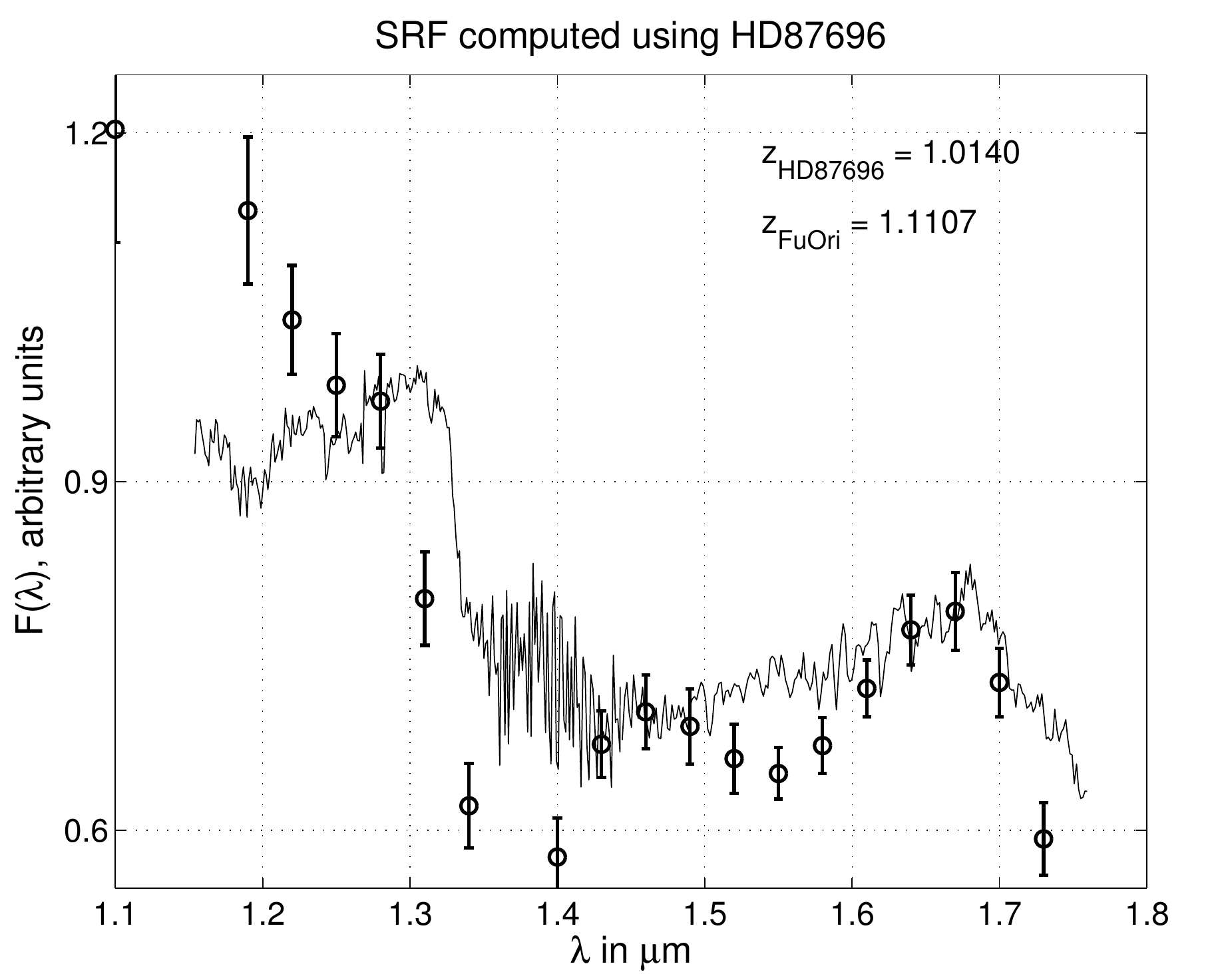}\\
\includegraphics[width=6cm]{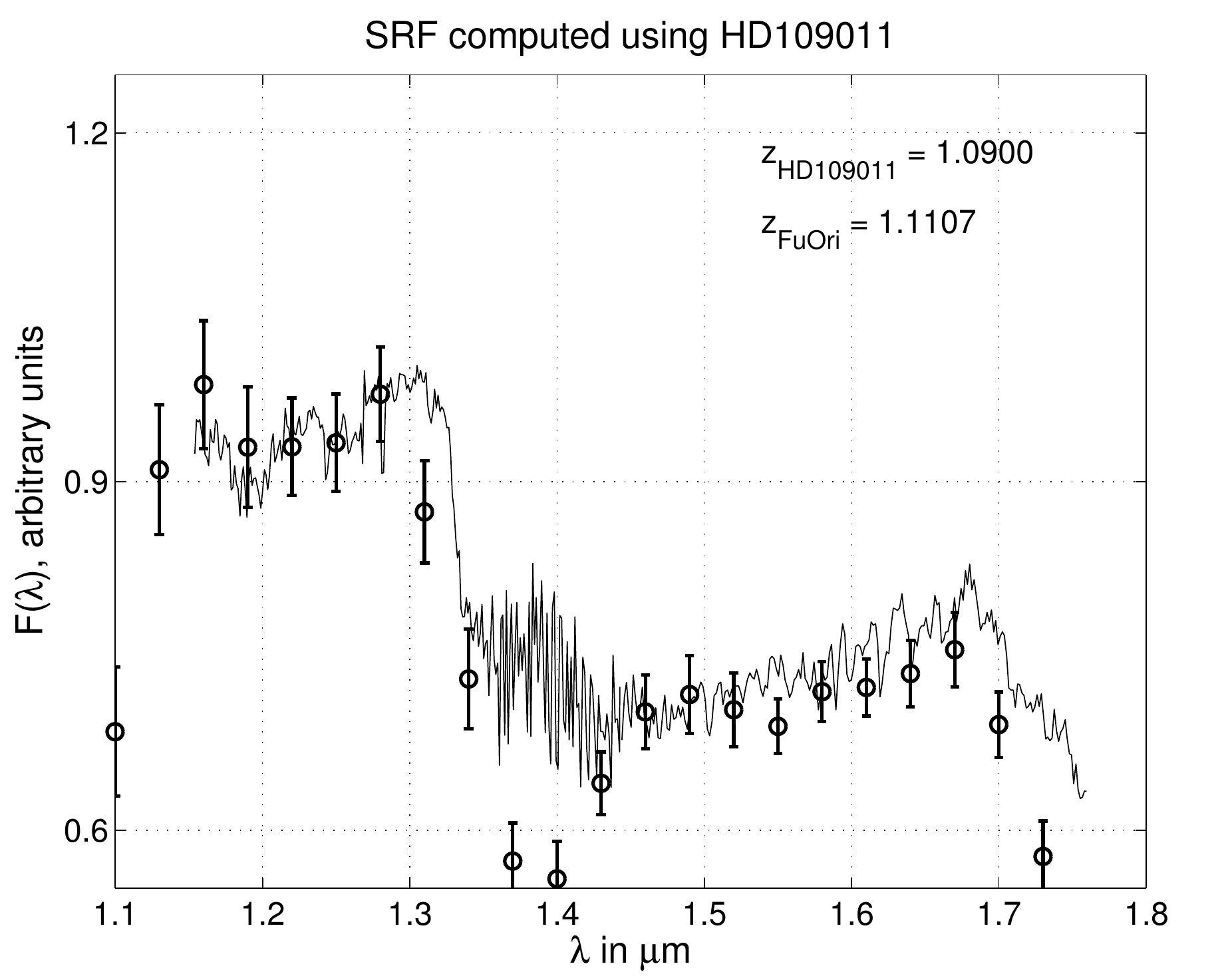}\\
 \includegraphics[width=6cm]{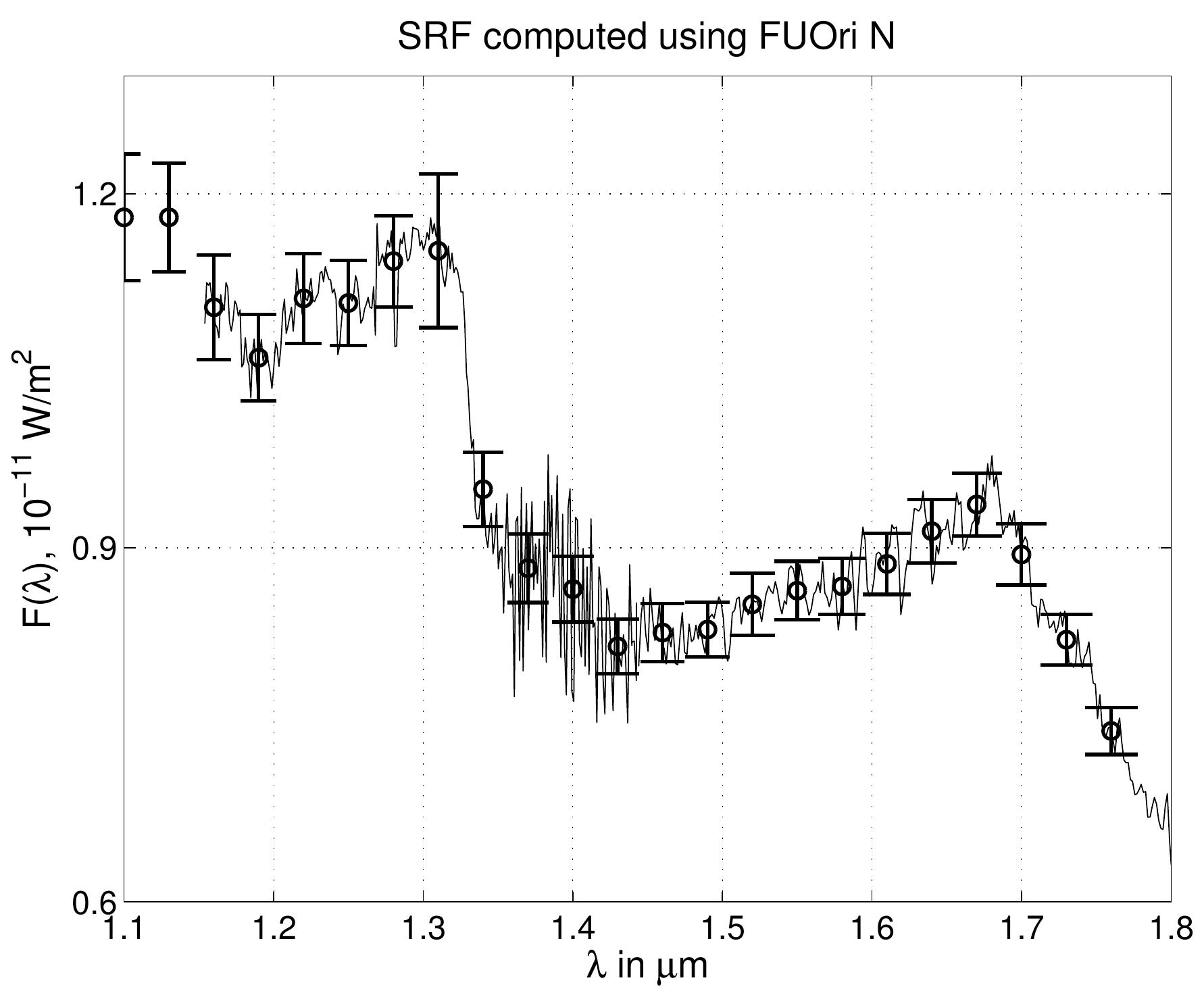}
\end{tabular}
\end{center}
\caption[Images]{Determination of the atmospheric and instrumental Spectral Response Function. {\em Circles}: P1640 spectrum of FU Ori N, obtained from non-coronagraphic images normalized using three different spectral response functions. {\em Solid Line}: IRTF spectrum of FU Ori N, from \citet{1996AJ....112.2184G} . {\em Top}: SRF calculated using HD104860, chi-squared difference with the IRTF spectrum $12.5$ {\em Second from top}: SRF calculated using HD87696, chi-squared difference with the IRTF spectrum $11.2$  {\em Second from bottom Left}: SRF calculated using HD109011, chi-squared difference with the IRTF spectrum $4.6$, note that most of the mismatch resides in the water absorption bands. The P1640 spectrum exhibits excellent agreement with the IRTF spectrum when the air-mass of the calibrator star is within $0.02$ from FU Ori. We thus conclude that the SED of FU Orionis has not varied since the 1994 \citet{1996AJ....112.2184G} and use the non coronagraphic images of FU Ori N to derive the SRF. {\em Bottom }: The final SFR is computed using FU Ori N as a reference star.}
\label{fig::SpectrumCalibrator}
\end{figure}

We call the Spectral Response Function (SRF) the wavelength dependent relationship between the spectrum of an astronomical source and its counterpart seen by the P1640 detector. In order to derive an accurate SRF, we use the non-coronagraphic images of FU Ori N that were acquired right before the coronagraphic observation of FU Ori S, combined with the published spectrum of FU Ori in \citet{1996AJ....112.2184G}. 
To do so we first need to establish that the J and H SED of FU Ori N has not varied since the 1994 epoch published by \citet{1996AJ....112.2184G}. We first derive three spectral response functions using three well characterized  stars 
($HD104860$ of spectral type F8V, $HD87696$ / A7V, and $HD109011$ / K2V) that were observed the same night as FU Orionis. We then compare our observed spectrum of FU Ori N, respectively normalized by each one of these three response functions.  Fig.~\ref{fig::SpectrumCalibrator} illustrates this comparison and shows excellent agreement for the calibrator within $0.02$ airmass of FU Orionis (HD109011). From this agreement we conclude that the $R\sim 45$ P1640 J and H SED slopes of FU Ori N have not varied since 1994, and thus combine the data from \citet{1996AJ....112.2184G} with our non-coronagraphic images to derive the final spectral response function we use to characterize FU Ori S, as illustrated in the bottom right panel of Fig.~\ref{fig::SpectrumCalibrator} . Note that the effect of using a calibrator at an airmass that differs from the FU Orionis system is most severe in the blue and red ends of the SED and in the water absorption band between J and H. Since the airmass change occurring during our coronagraphic observing sequence is $0.02$ (i.e. identical to the difference between the FU Ori and the HD109011 exposures, bottom left on Fig.~\ref{fig::SpectrumCalibrator}) it is very difficult to constrain the spectral calibration uncertainties in these regions of SED. We thus chose to discard these points for our analysis of the SED of FU Ori S. 

\subsection{Extraction of the spectrum of FU Ori S}
\label{sec::SpectralExtraction}
\begin{figure}
\begin{center}
\begin{tabular}{c}
\includegraphics[width=8cm]{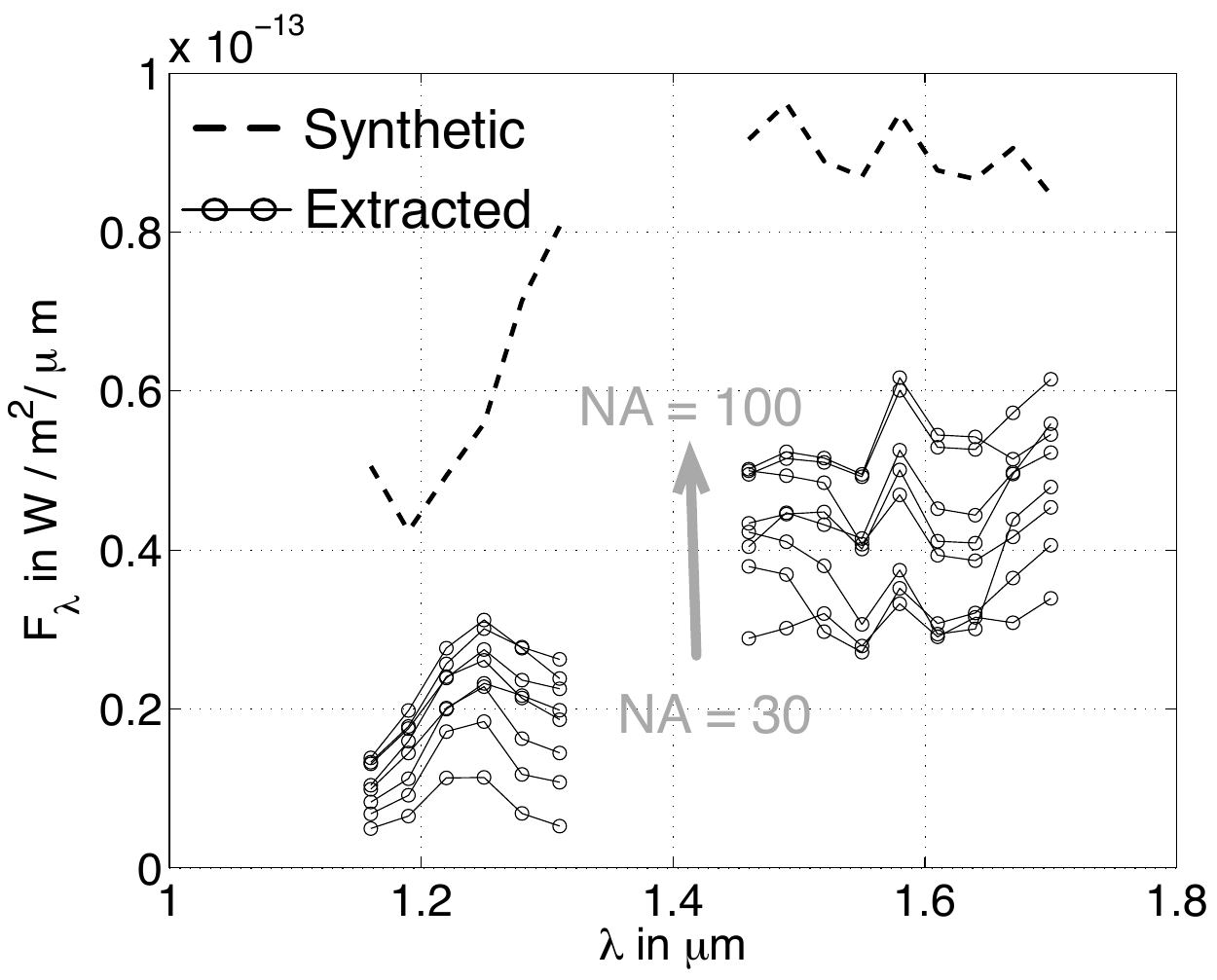}\\
\includegraphics[width=8cm]{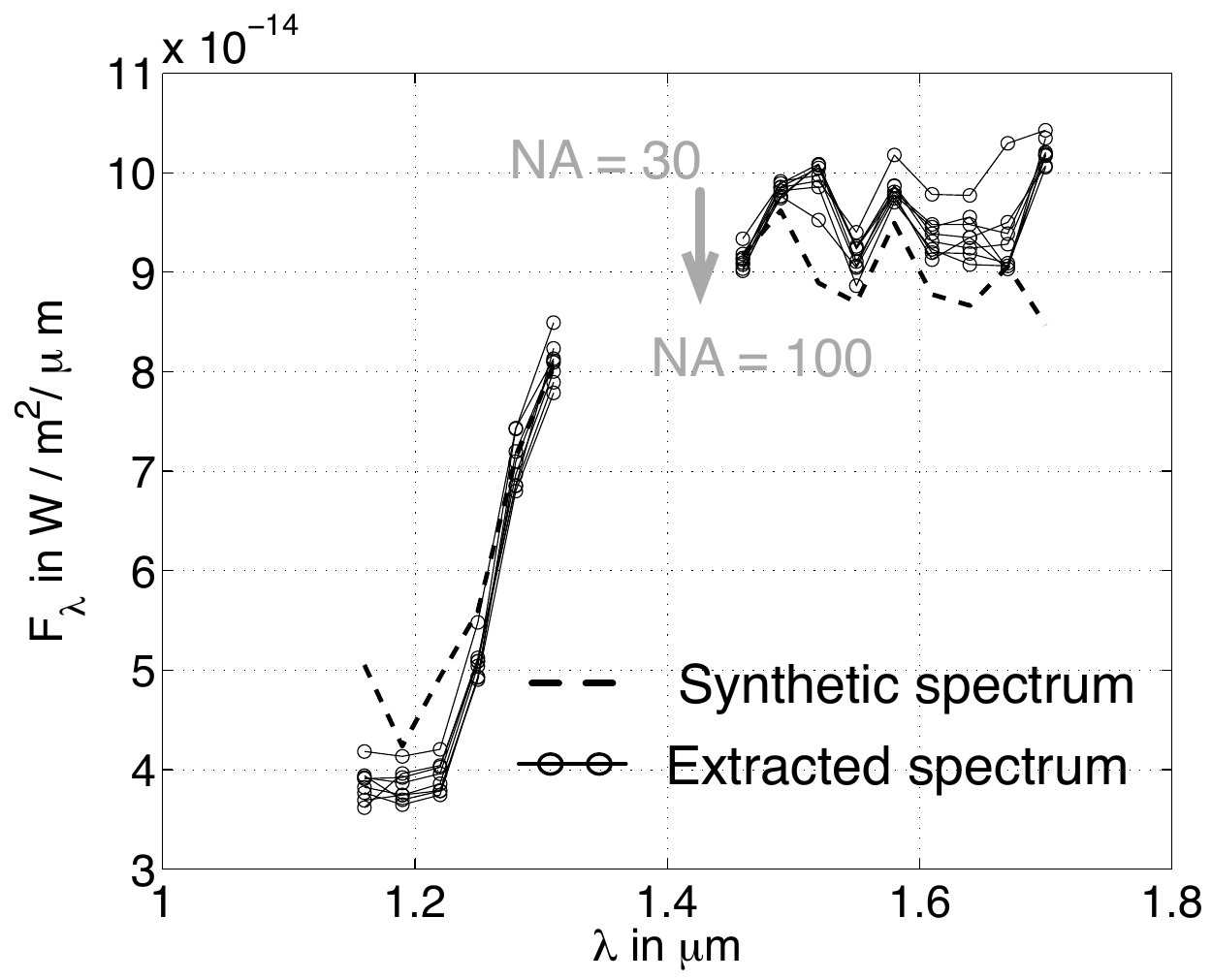}
\end{tabular}
\end{center}
\caption[Images]{Quantification of LOCI and d-LOCI bias using synthetic companions. The spectra are shown before calibration by the Spectral response Function. In both panels the dashed line corresponds to the synthetic companion injected in the coronagraphic PSFs. This ``zeroth order'' spectrum was obtained using aperture photometry on preliminary LOCI reduced images, and renormalized to match the band averaged photometry of the raw data. The top panel shows the synthetic spectrum extracted after LOCI: while the images exhibit the high SNR illustrated on Fig.~\ref{fig::ImagesFuOri}, the  spectro-photometric signal clearly exhibits an algorithmic flux depletion. Moreover this bias depends upon the reduction parameters. We alleviate this problem using a d-LOCI approach, bottom panel, which only exhibits a small bias over a wide range of  reduction parameters.}
\label{fig::ExtractedBias}
\end{figure}

To mitigate the contamination of the spectro-photometry of FUOri S by residual speckles, we first seek to calibrate these residual quasi-static optical artifacts using an aggressive PSF subtraction algorithm, the Locally Optimized Combination of Images approach \citep{JustinSpeckle, 2007ApJ...660..770L}. For a given location in the image, LOCI creates a synthetic reference PSF, a weighted sum of images within a collection of reference frames, based on a least-square fit of the neighboring speckles. In order to preserve as much flux from the companion as possible, the least squares fit is calculated using a large  ``optimization'' region of the image, with area expressed as $N_{A}$ PSF cores, while the actual subtraction is carried out in a smaller ``subtraction'' zone. As discussed in \citet{LaurentSpeckle}, direct photometry on images processed using classical LOCI implementations, such as the reduced images in the bottom panel of Fig.~\ref{fig::ImagesFuOri} for instance, can produce biases in the spectro-photometry. We quantified this effect, by extracting a ``zeroth-order'' spectrum of FU Ori S from the reduced data cubes of Fig.~\ref{fig::ImagesFuOri}, and used this spectrum to inject a set of synthetic companions in our dataset. 
The top panel of Fig.~\ref{fig::ExtractedBias} shows, for a typical azimuthal orientation, the injected synthetic spectrum and a series of extracted spectra after LOCI reduction.  Each extracted spectrum was estimated using a reduction with a given area of least-squares minimization, $N_A$. Using such a reduction strategy leads to underestimating the flux of the companion by a factor of two to three. Moreover this bias is wavelength dependent. As presented in \citet{LaurentSpeckle} this flux depletion is a combination of a least squares bias common to all LOCI implementations (that can be identified as a ``grey'' gain that does not alter neither spectral features nor the SED slope), and a spectral cross talk term, specific to integral field spectrographs. 

We circumvented this problem by using the ``damped LOCI'' approach introduced in \citet{LaurentSpeckle}. d-LOCI relies on a least-squares approach similar to the one described above but adds a supplemental penalty term, that scales with the flux of the discovered companion, in the underlying quadratic cost function. This preserves the flux from faint sources and leads to un-biased SED estimated even in the case of a companion buried under speckles. The results with synthetic companions are shown in the bottom panel of Fig.~\ref{fig::ExtractedBias}. Injected and extracted spectra agree very well for a wide range of d-LOCI parameters. We further explored potential reduction biases by varying  other algorithm parameters that scale with the cross-spectral channels contamination. d-LOCI reductions yield consistent results for a wide range of parameters, with a bias smaller than 5 percent. We thus conclude that our final spectrum is unbiased at the $\sim 5$ percent level. 
 
\begin{figure}
\begin{center}
\includegraphics[width=8cm]{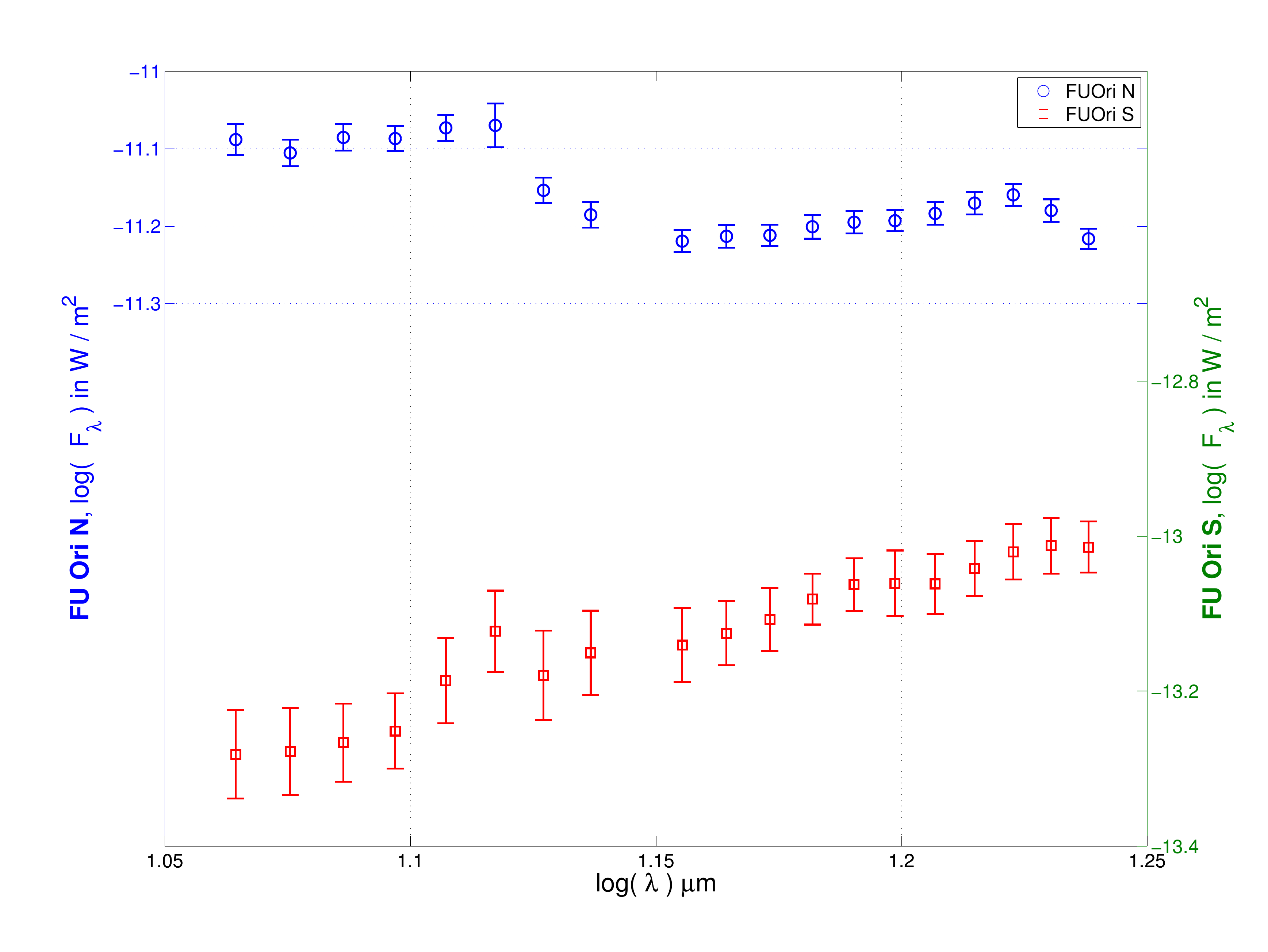}
\end{center}
\caption[SED]{J and H band Spectral Energy Distribution of the FU Orionis system. {\em Blue Circles}: P1640 spectrum of the Northern component. {\em Red Squares}: P1640 spectro-photometry of the Southern component.}
\label{fig::CompareSEDs}
\end{figure}

In Figure \ref{fig::CompareSEDs} we present our final spectrum for FU Ori S and, for comparison, that of FU Ori N.  We derived the error bars as the root mean square of the sum of three separate terms: the error on the spectral response function (e.g. photometric scatter in non coronagraphic images), the scatter of bias estimated on synthetic companions over the ensemble of LOCI parameters explored (smaller than $5$ percent), and the photometric scatter in the reduced images. The spectrum of FU Ori S is rising to the red throughout the J and H bands and appears to be featureless.


\section{Physical association of the two components of the FU Orionis system}

\subsection{Proper Motion of FU Orionis}

The proper motion of FU Ori N is reported both in the Carlsberg Meridian Catalog \citet{2002A&A...395..347E} \footnote{http://archive.ast.cam.ac.uk/camc/}, as $pm RA = 0.116'' /$ year $\pm 0.104$,  $pm DEC = 0.0864'' /$ year $ \pm 0.0115$, and  Proper Motion Extended Catalog (PPMXL),  as  $pm RA = 0.0144''/$ year $\pm 0.0108$ and $pm DEC = 0.0721''/$ year $\pm 0.0108'$. Although the proper motion estimates in right ascension have large percentage error, the proper motion in declination is highly significant. If we adopt a distance of $450$ pc to FU Orionis, these two estimates yield  surprisingly large transverse velocities, $183$ km/s and $139$ km/s respectively. Tracing back the trajectory of FU Orionis to $\sim 1 \; M_{yrs}$ using these values places this object in the T-Tauri star forming M78. While there exists a multitude of observational evidence for the young age of FU Orionis, its location, seemingly isolated from any star forming cluster, can be seen as somewhat puzzling. This large proper motion could explain this paradox, and in this scenario FU Ori could be a runaway star from M78. However the large ejection velocity required to explain this scenario can only be the result of an unlikely very close encounter early in the history of FU Orionis. Moreover such transverse velocities do not seem compatible with  the radial velocity of FU Ori N measured by \citet{2003ApJ...595..384H}, which is consistent with the molecular cloud velocity. Alternatively these proper motion estimates, based on optical images,  could be biased by the extended optical nebulae surrounding FU Orionis. We further explored this issue and derived proper motion using astrometric estimates based solely on infrared observations, since at longer wavelength FU Ori is a true point source. To do so, we take advantage of the $\sim 10$ years baseline between 2MASS and WISE observations. Using the 2MASS point source catalog and the WISE preliminary catalog, we obtain  $pm RA = 0.033''/$ year $\pm 0.14$,  $pm DEC = 0.025''/$ year $\pm 0.053$. While the mean value of this proper motion estimate yields a smaller transverse velocity of  $56$ km/s, the large declination uncertainty in the preliminary WISE point source catalog, due to a systematic bias in WISE astrometric estimates that will be corrected in a future release, prevents us from making a definitive conclusion regarding the actual true proper motion of FU Orionis. For the purpose of demonstrating that FU Ori North and South are co-moving, we will use the smallest of these three proper motion values, the 2MASS to WISE infrared to infrared estimate. 

\subsection{Physical association of FU Ori S and FU Ori N}

As the one year time baseline between the epochs presented in \citet{2004ApJ...608L..65R} and \citet{2004ApJ...601L..83W} was not sufficiently large to establish or rule out physical association between the FU Ori North and South, we explore this question using  the 2009 epoch from P1640. In our data we detect FU Ori S at a separation of $0.491'' \pm  0.007$ and a position angle of $ 161.2^{\circ} \pm  1.1$ with respect to FU Ori N. The astrometric pupil plane grid, necessary to carry out astrometry on occulted coronagraphic images \citep{2006ApJ...647..620S,2010ApJ...712..421H,2011ApJ...726..104H}, was not in the optical path during the observing sequence of FU Orionis.  Astrometric estimate were thus carried out using the un-occulted reddest images, where the companion is visible, albeit at very low SNR. Each of the five images chosen in the H band was analysed with the program FITSTARS; which uses an iterative blind-deconvolution that fits the location of delta functions and their relative intensity to the data. The  FITSTARS program was presented in \cite{tenBrummelaar1996, tenBrummelaar2000}. After throwing out any data that failed to converage to a physical solution, the position angle and separation was computed by a weighted average of the individual values. The weights were set equal to the inverse of the RMS residual of the fit; a standard output of the FITSTARS program. The error bars for position angle and separation were set equal to the weighted standard deviation of the results.

\begin{figure}
\begin{center}
\includegraphics[width=9cm]{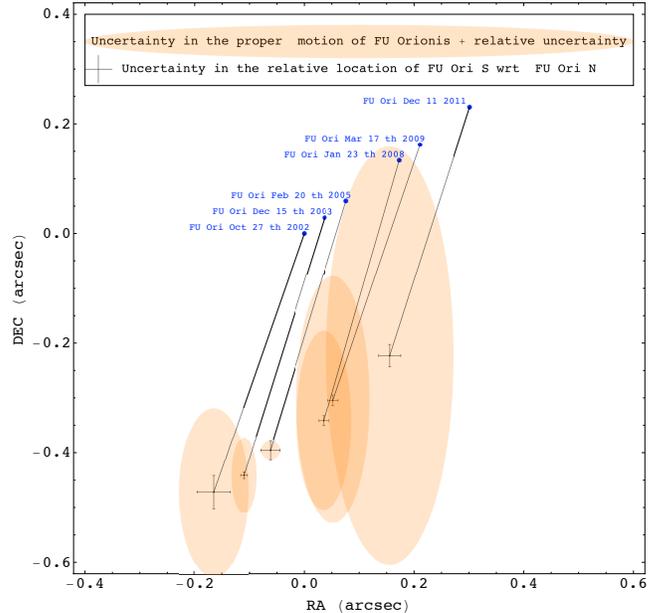}
\end{center}
\caption[ProperMotion]{Proper motion of the FU Orionis system. The blue dots represent the location of FU Ori N across the five epochs considered. The star motion was calculated using the astrometric difference of the WISE and 2MASs epochs. The uncertainties on this motion, have been cumulated with the uncertainty in the relative position of FU Ori S with respect to FU Ori N to estimate the regions of confidence of the on-sky location of FU Ori S (orange ellipses). For clarity, we arbitrarily used the 2005 epoch as the origin of time for the calculation of the uncertainties in proper motion and the 2002 epoch as the zero point of our diagram. Since the regions of confidence between the 2002 and the 2003 epochs overlap, common proper motion could not be established at the time. However the data points introduced in this paper provide much greater temporal leverage  and demonstrate that FU Ori N and FU Ori S are physically associated.}
\label{fig::FUOriProperMotion}
\end{figure}

To increase confidence in our astrometric point,  we included in our common proper motion analysis a second P1640 epoch, obtained in 2011 with the astrometric calibration grid in the optical train. We also complemented our observation using the  2005 epoch obtained as part of the interferometric survey presented in \citet{2009ApJ...700..491M}, and re-processed with particular care to extract the flux from FU Ori S. The result of our analysis is summarized in Figure \ref{fig::FUOriProperMotion}, which includes points previously published,  where both the uncertainties in the relative position of both components and the proper motion of FUOri N have been combined as an uncertainty in the position of FU Ori S. The configuration of this system is somewhat cumbersome as the largest source of uncertainty is the lack of a well constrained proper motion estimate for FU Ori N. Even in the presence of such large uncertainty Figure \ref{fig::FUOriProperMotion} shows that the combination the five epochs clearly establishes that FU Ori N and FU Ori S are co-moving. We thus conclude that these two objects are physically associated.  The relative position of the binary pair across the six epochs in Figure \ref{fig::FUOriProperMotion} is given in Table \ref{tab::FuAstrometry}. Under the assumption of a face on orbit and a $0.4 \; M_{\bigodot}$ mass for FU Ori N, the largest orbital motion detectable over the 9 year baseline provided by these points is $26$ micro-arcseconds in the NS direction assuming that FU Ori S is massless and $45$ micro-arcseconds in the NS direction if the mass of FU Ori S is twice the mass of FU Ori N. In any case, the astrometric uncertainties in Table \ref{tab::FuAstrometry} are much larger than these optimistic projected relative motion, and we conclude that the scatter in these values is not representative of any significant orbital motion. 

\section{Reconstruction of the 0.8-10 $\mu$m Spectral Energy Distribution of FU Ori S}

\subsection{Consistency with published J and H band observations of FU Ori S}

We first focused on comparing the P1640 band averaged photometry with the estimates obtained by \citet{2004ApJ...608L..65R}. By integrating the data points in Fig.~\ref{fig::CompareSEDs} over the J and H bandpasses, we found delta magnitudes between FU Ori N and FU Ori S of $\Delta m_{J} = 5.6$ mag and $\Delta m_{H} = 4.9$ mag that are significantly larger than the values estimated by Reipurth \& Aspin, respectively $4.5$ mag and $4.3$ mag. 

\begin{figure}
\begin{center}
\includegraphics[width=10cm]{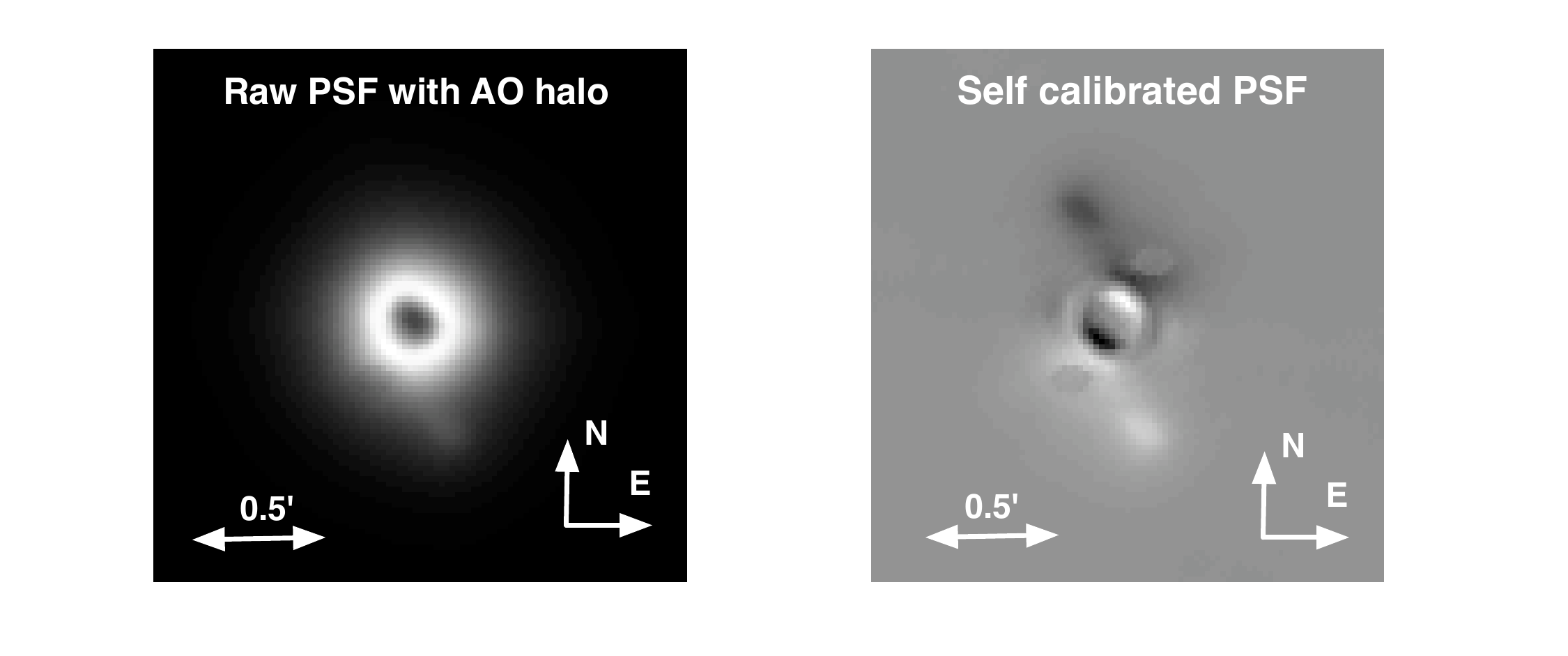}
\end{center}
\caption[Images]{In an effort to reconcile the P1640 photometry with the IRCS data we reprocessed the 2003 epoch published in \cite{2004ApJ...608L..65R}. {\bf Left:} raw IRSC PSF.  FU Ori S is embedded within the adaptive optics halo. {\bf Right:} Self-calibrated AO PSF, obtained by self subtracting each frame with its the centro-symmetric image. In the raw image the halo contaminates the photometry of FU Ori S and our gaussian fitting analysis yield estimates similar to \cite{2004ApJ...608L..65R}. The same analysis on the Self-calibrated AO PSF, where the halo contamination has been mitigated, yields photometric estimates consistent with values obtained on P1640 data.}
\label{fig::IRCSImages}
\end{figure}

We thus closely inspected the imaging data from \citet{2004ApJ...608L..65R}. These images were obtained using the Infrared Camera for Surveys at the Subaru Telescope and are publicly available on the SMOKA archival system \footnote{http://smoka.nao.ac.jp/}. They consist  of a series of images  in which the brightest star is saturated, and a series of photometric calibration frames where the brightest star is attenuated by a $1/100$ neutral density filter. The left panel of Fig.~\ref{fig::IRCSImages} illustrates the raw J band IRCS data; the FU Ori S component is detected at $SNR \sim 3$. It is heavily embedded in the residual adaptive optics halo. We can reproduce the delta magnitudes reported in \citet{2004ApJ...608L..65R} by carrying out a gaussian photometric fit on these images. However, in an effort to further separate the photometric contributions of the AO halo and FU Ori S itself, we proceeded to subtract each exposure with a centro-symmetric image of itself (right panel of Fig.~\ref{fig::IRCSImages}, similar to Fig.~1 in \citet{2004ApJ...608L..65R}) before carrying out the photometric estimation. This process led to substantially larger estimates of the magnitude difference,  $\Delta m_{J} = 5.3$ mag and $\Delta m_{H} = 5$ mag. These values are better matched to our P1640 observations, even consistent within generous error bars. We then carried out the same procedure for the IRCS K' and L' band images of FU Ori S. Since the AO halo decreases with wavelength, these long wavelength images exhibit $SNR >10$ and thus we found that both methods outlined in Fig.~\ref{fig::IRCSImages} yield estimates consistent with the values published by \citet{2004ApJ...608L..65R} . The results of our photometric analysis are summarized in Table \ref{tab::FuPhotometry}. Note that the main conclusion of \citet{2004ApJ...608L..65R}, namely that FU Ori S is a young star very likely to be associated with FU Ori N, relied on red J-H colors. It is thus not impacted by our significantly revised photometric estimates which indicate an even redder object.


\citet{2012AJ....143...55B} recently reported high spectral resolution J H and K band ($R \sim 3000$) spectra of FU Ori S, obtained with the Gemini NIFS integral field spectrograph. They reported a positive SED slope in J and a negative slope in H, while our estimates, Fig.~\ref{fig::CompareSEDs}, yield positive slopes in both bands. Although this could be the result of extreme near infrared color variability of FU Ori S, we have argued against such varaibility for FU Ori N in section 2.2, and indeed there may be a technical explanation for the difference in spectral slope between the \citet{2012AJ....143...55B}  and our observations. While P1640 and NIFS are both integral field spectrographs, their design tradeoffs are radically different. NIFS prioritizes spectral resolution ($R \sim 3000$) over spatial resolution ($ \sim 10$ angular resolution units per spaxel at the shortest wavelength). P1640 was designed specifically to explore an orthogonal parameter space, with a fine spatial resolution in order to mitigate speckle contamination ($R \sim 45$, $>2$ spaxels per unit of angular resolution at the shortest wavelength). 

In order to investigate the H band slope discrepancy, we carried out aperture photometry estimates on our P1640 data without any speckle suppression. This led to an SED estimate similar to the synthetic spectrum shown Fig.~\ref{fig::ExtractedBias}, which mimics the SED profile in \citet{2012AJ....143...55B}. We thus conclude that the discrepancy in H band slope could be due to mid-spatial frequency speckle contamination in NIFS, that can not be calibrated because of the large angular extent of the NIFS spaxels.  Alternatively it could be the result of a product near infrared variability of FU Ori S and require further monitoring of this system. Since the characterization of FU Ori S in \citet{2012AJ....143...55B} is based on line diagnostics that take advantage of the high spectral resolution, revising their estimated SED slope would not alter their conclusions. 

\subsection{Assembled Spectral Energy Distribution of FU Ori S}

We complemented our J and H SED derived from P1640 data with published measurements spanning $0.8 \; \mu m $ to $10 \;\mu m$. The  $0.8 \; \mu m $ point was obtained using the delta magnitude of $\Delta m_{0.8} = 3.96 \pm 0.28.$ reported from speckle imaging by \citet{2008AstBu..63..357K}. It was placed within the overall SED of FU Ori S using an extrapolation to $0.8 \; \mu m $ of the SED of FU Ori N published in \citet{1996ARA&A..34..207H}. We used the K band spectrum published by \citet{2004ApJ...608L..65R}, scaled to match the estimated K photometry obtained using our  analysis of the IRCS data. Comparison of the \citet{2004ApJ...608L..65R}  and the Beck \& Aspin spectral slopes through the K band shows that they are similar, if not identical, outside of the CO bandhead region.

The IRCS L band photometry we derived was also incorporated in our SED reconstruction. Finally the $10.2$ micron $N $ band point was obtained using a non-redundant aperture configuration of the Keck segment tilting experiment, as reported in \citet{2009ApJ...700..491M}, but with special attention in the data reduction for our purposes to the presence of two point sources.  FU OriN and FU OriS are separated by $475 \pm 10$ mas and have a position angle of $163 \pm 2$ deg; the N-band flux ratio is $\sim10\pm 1.5$.  Neither component of the binary is itself spatially resolved in the 10 $\mu$m data.

To the best of our knowledge there is no other ultraviolet, optical, infrared, or sub-mm data that spatially resolves the pair and therefore could be incorporated into our analysis.

\section{Analysis of the Spectrum and Spectral Energy Distribution of FU Orionis S}

\begin{figure}
\begin{center}
\includegraphics[width=7cm]{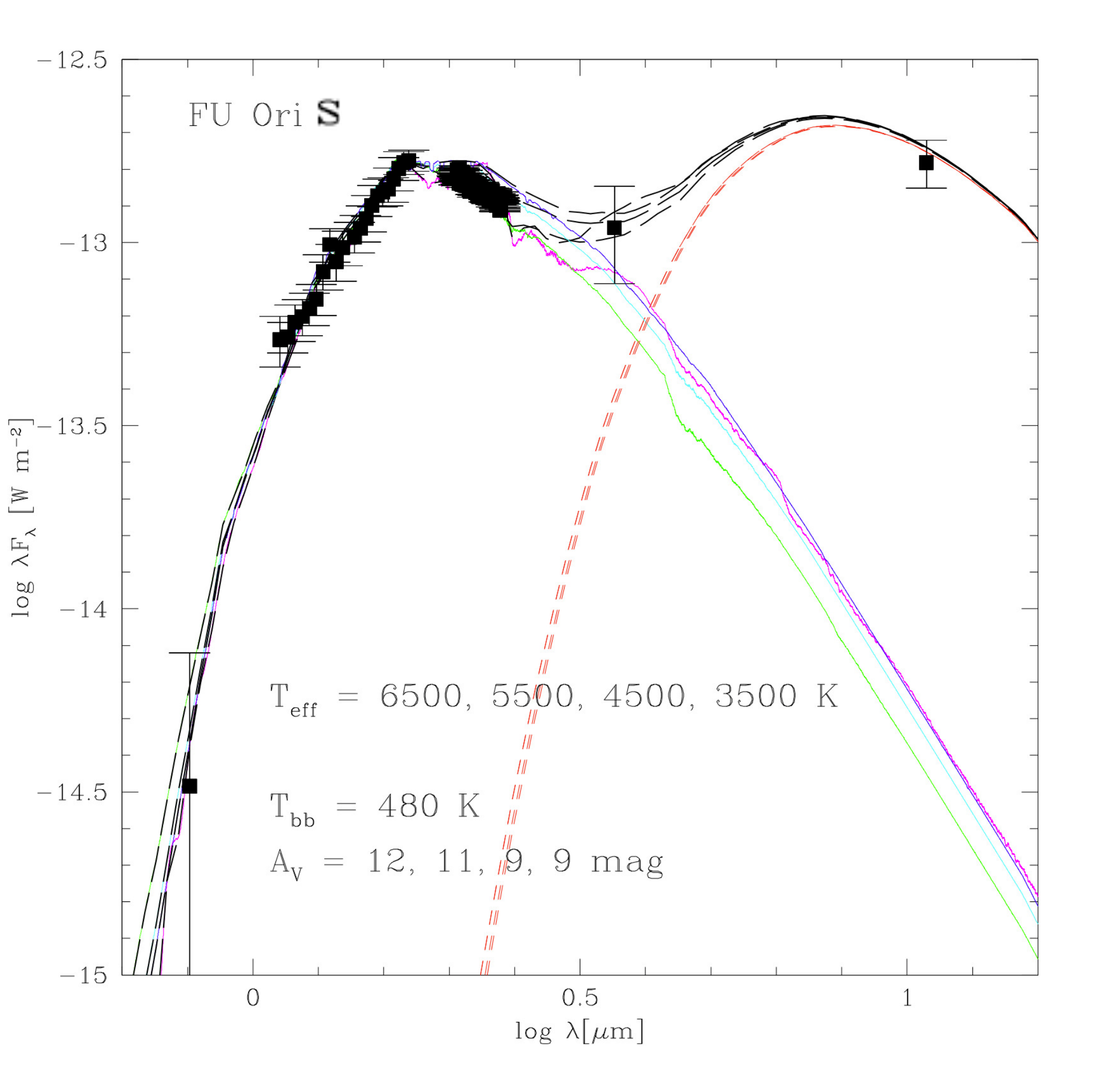}
\end{center}
\caption[SEDModel]{Overall Spectral Energy Distribution of the FU Orionis system with best fit models. The N-band point is well above any conceivable reddened stellar photosphere, and must therefore be due to circumstellar dust around FU Ori S. The colored curves are NextGen model atmospheres at the representative effective temperature and extinction pairs $T_{eff}=6500$, $A_V=12$ (blue), $T_{eff}=5500$, $A_V=11$ (cyan), $T_{eff}=4500$, $A_V=9$ (green), $T_{eff}=3500$, $A_{V}=9$ (magenta) all with surface gravity $log\; g = 4.5$.  The red line is a 450 K blackbody with the same extinction as the matched photosphere.   The sum of the photospheric and excess contributions are represented by the black lines, which are normalized to the data at the P1640 $1.7 um$ point.The L-band and N-band points can be fit adequately by adding  a single reddened blackbody of temperature $\sim 480$ K to reddened photospheres.}
\label{fig::grid}
\end{figure}

Considering first the spectral line information available in our $R\sim 45$ P1640 data, the J- and H-band spectrum lacks any absorption features.  This constrains the temperature to be warmer than $\sim$4000 K since at cooler temperatures molecular absorption from CO and $H_2O$ would become apparent even at the low spectral resolution of P1640 and in the presence of high extinction. Meanwhile, at warmer temperature narrow atomic features may be present, but would not be visible at the P1640 resolution. The published $R\sim 1500$ K-band spectrum from \citet{2004ApJ...608L..65R} hints at the presence of NaI at 2.206 um and perhaps CaI at 2.26 $\mu$m that the authors invoke to advocate for a temperature corresponding to a G or K spectral type.  This is consistent with their otherwise featureless K-band spectrum that implies a temperature warmer than $\sim$4500 K based on the absence of CO absorption, and cooler than $\sim$7000 K due to absence of BrG absorption (though an actively accreting  young star could have both CO and BrG absorption filled in to the continum level without these lines appearing explicitly in emission).  Higher signal-to-noise and higher spectral resolution data K-band data recently published by \citet{2012AJ....143...55B} more clearly exhibit the photospheric absorption features in FU Ori S, and suggest a spectral type of K5.  Notably, these authors find both the CO bandhead and BrG in {\it emission}, whereas the earlier \citet{2004ApJ...608L..65R} spectrum showed no emission in these regions. It is somewhat unusual for a young stellar object to have both clearly present photospheric absorption features and strong CO emission.    

The overall SED of FU Ori S from R- to J- to H-band is quite red, see Fig.~\ref{fig::grid}.  Considering, in addition, the K-, L-, and N-bands, the SED is also appreciated as broader than a photospheric spectrum.  An analysis of the available SED was carried out using both blackbody fitting and comparison to stellar atmospheres in order to derive rough temperature and extinction estimates. We find that the N-band point is well above any conceivable reddened stellar photosphere, and must therefore be due to circumstellar dust around FU Ori S.  The L-band point is also likely in excess of the stellar photosphere though for very cool, highly reddened atmospheres, an L-band excess is not required. Without a more fully populated spectral energy distribution the disk can not be characterized in terms of mass or size, but only in terms of the hottest dust closest to the photosphere.  The L-band and N-band points can be fit adequately by adding to the reddened photosphere described below a single reddened blackbody of temperature $450-480 \;K$ . The de-redened luminosity of FU Ori N is dominated by its hot inner disk and has been estimated at $226 L_{\odot}$ \citep{2007ApJ...669..483Z}. When combining effective temperature estimates based on our reconstructed SED (see below)  with a stellar radius of $\sim 1 R_{\odot}$ we find that  de-reddened luminosity of FU Ori N is $\sim 0.6 L_{\odot}$. As a consequence, assuming a distance of $225$ AU between the two components, we find that for separations $<10$ AU the dusty material around FU Ori S is mostly heated by FU Ori S and that further away heating from FU Ori N becomes dominant. As the temperature from this dust, as per our SED estimate, is $~\sim 480 K$, then it is mostly likely lying within the inner AU of the southern component and is most likely heated by FU Ori S.

\begin{figure}
\begin{tabular}{c}
\includegraphics[width=7cm]{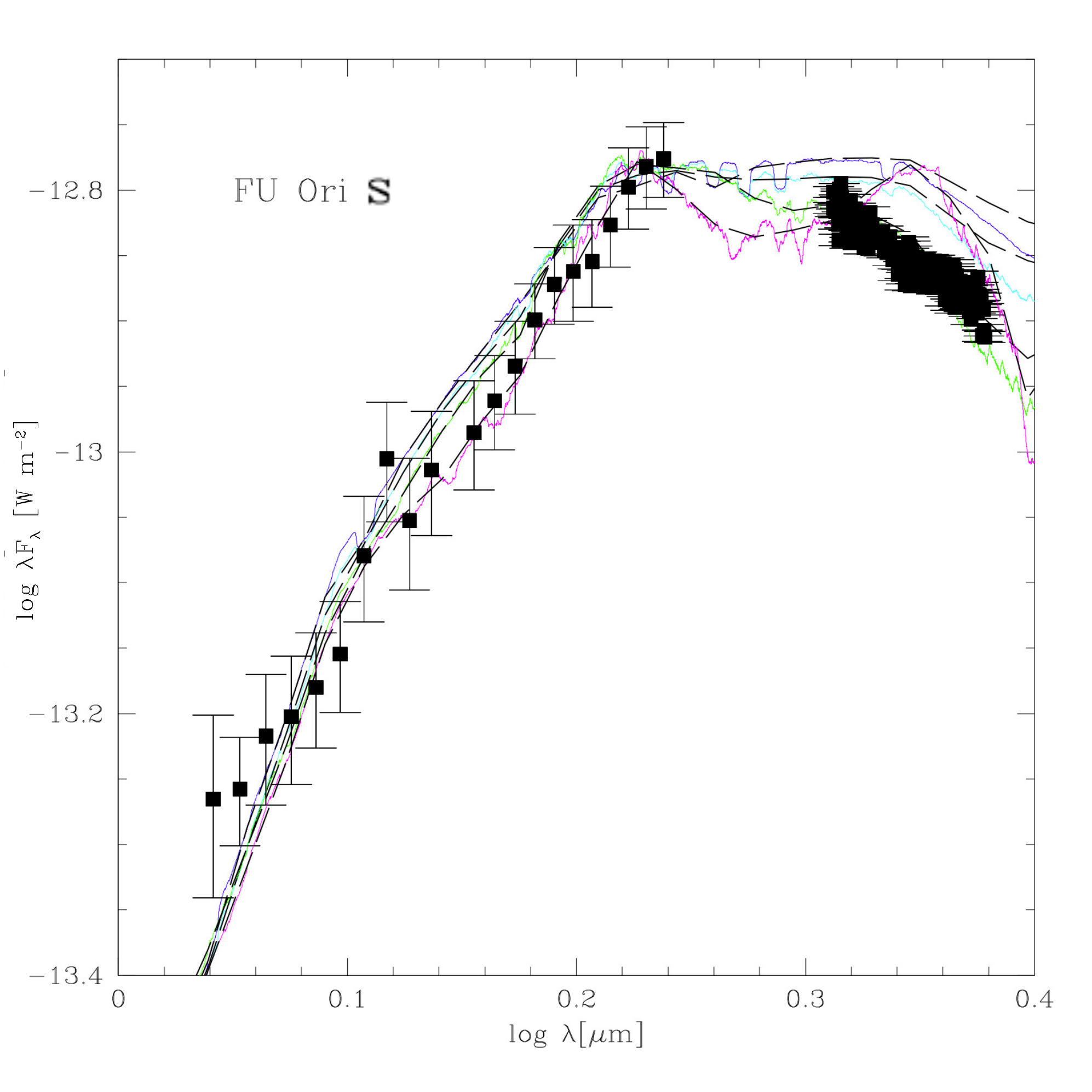}\\
\includegraphics[width=8cm]{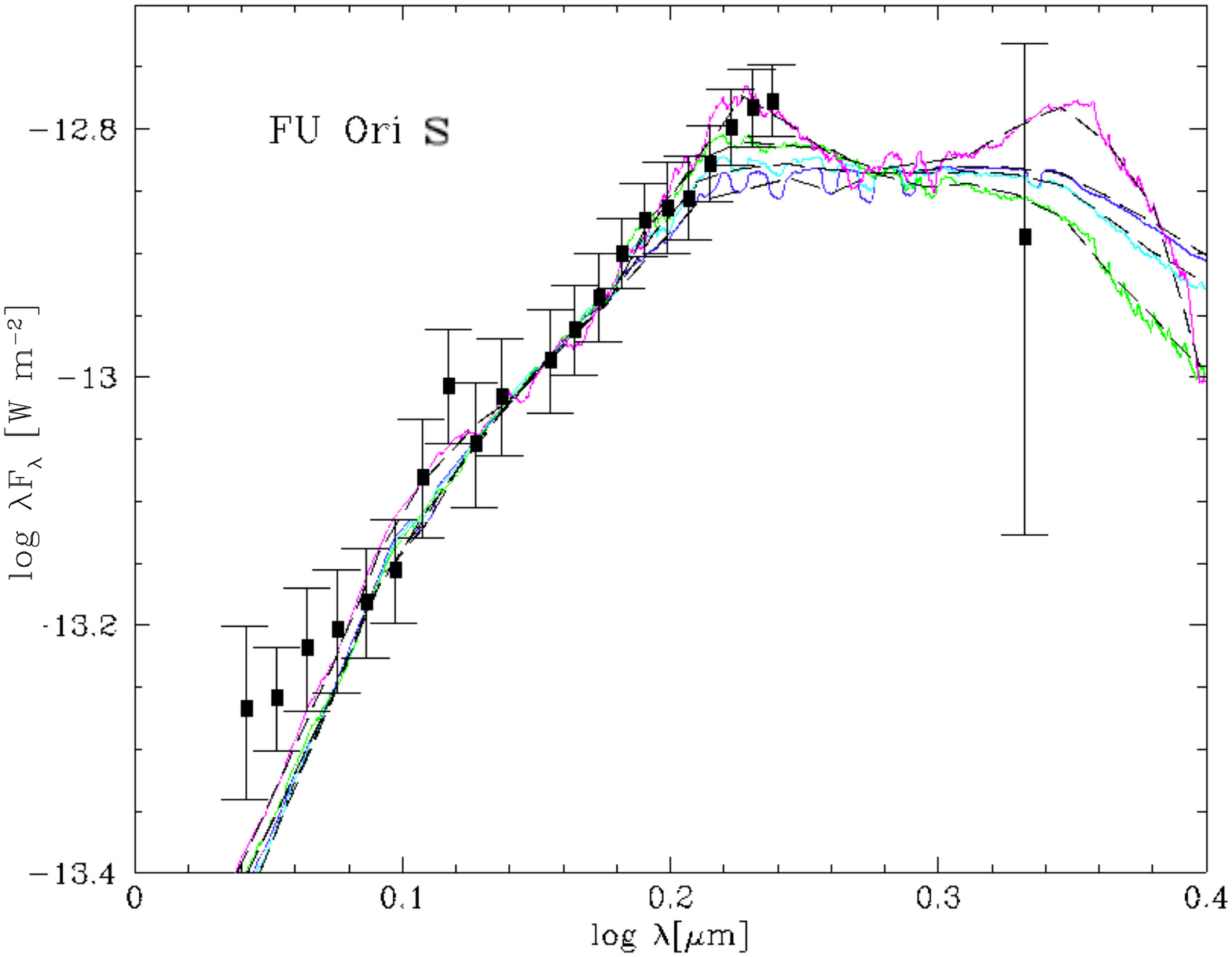}
\end{tabular}
\caption[SEDModelZoom]{Zoom on the region of the SED probed by P1640. Same colors as \ref{fig::grid}.{\em Top}: all the models are scaled to match the P1640 spectrum at $1.65 \; \mu$m. {\em Bottom}: all the models are scaled to match the P1640 spectrum at $1.4 \; \mu$m. Agreement between data and models depends on the wavelength at which the models are normalized. We find the best-fit for an atmosphere with $T=4000 \; K$ and $A_V=8$, normalized at either the short end of H or the long end of J.}
\label{fig::gridZoom}
\end{figure}

Moving blueward, the origin of the K-band flux is more ambiguous.  The shape of both published K-band spectra (excluding the CO bandhead region which has different slope in the \citet{2004ApJ...608L..65R}  vs. the \citet{2012AJ....143...55B} spectra, perhaps indicating the onset of CO emission) is consistent with an un-reddened blackbody of roughly 2200-2500K.  The shape of the JH region P1640 data, however, requires a much more extincted spectrum, for example Av=7 mag for the same 2500 K blackbody. This model does not then fit the published K-band spectral shape though it does pass through the mean K-band flux.  Considering alternate blackbodies, Av=10 mag is required for a $\sim$5000K temperature blackbody to fit the JH spectral region, but again this does not fit the K-band spectral shape very well, though does still match the mean K-band flux.  Model atmosphere fitting in the JHK wavelength regime introduces sensitivity to spectral signatures that would arise in cool atmospheres.   For this purpose we used the NextGen \citep{2001ApJ...556..357A} models released in 2005\footnote{http://phoenix.ens-lyon.fr/Grids/NextGen/SPECTRA/}.  As seen on Fig.~\ref{fig::gridZoom}, there is a general trade-off between extinction and temperature through the P1640 J-band and H-band region, with $T = 6500, 5500, 4500$, and $3500 K$ models best fit with $A_V = 12, 11, 9$ and $9$ mag, respectively, as illustrated. Cooler temperatures exhibit broad molecular absorption features throughout the J and H bands, which as noted above are not seen. The relatively featureless K-band spectrum additionally rules out the $3500 K$ model for the same reason that as even cooler models are ruled out from the J-and H-band data alone.  The ability of a model with given temperature and extinction to match both the steep positive slope through the J- and H-band region and the negative slope through the K-band, depends on the wavelength at which the model is normalized to the data.  For example, a normalization at the far red end of the H-band permits a wider range of temperature plus extinction models to pass within the error bars through the R-band photometry and the J- and H-band spectrophotometry, as well as to come close to matching the K-band slope, than can be fit to the same $\chi^2$-based confidence level when normalizing at shorter wavelengths.  Shorter wavelength H-band or even J-band normalization allows the red end of the H-band as well as the K-band to serve as a discriminant between models, see Fig.~\ref{fig::gridZoom}. Overall, we find the best-fit for an atmosphere with $T=4000 K$ and $A_V=8$, normalized at either the short end of H or the long end of J. This effective temperature is consistent with \citet{2012AJ....143...55B,2004ApJ...608L..65R}. our P1640 observations provide tighter constraints on this quantity along with an estimate of the extinction in the line of sight of FU Ori S.  

\section{New Results on FU Ori S in the Context of FU Ori N}

In this paper we have established that FU Ori S is physically related to FU Ori N and the pair forms a true binary system. The P1640 astrometry over 2 epochs combined with literature information establishes common proper motion of the pair. Our spatially resolved spectrophotometry in JH in combination with the limited amount of data at other wavelengths that spatially resolves the binary enables for the first time an assessment of the N/S pair of young stars in the FU Orionis system. The northern component is a long recognized FU Ori star that is visible at optical wavelengths and well-studied in the context of the accretion disk-dominated paradigm for FU Ori objects, the class for which it is the prototype. Previous modelling work has estimated that the underlying star is a 0.3 to 0.5 $M_\odot$ star behind $A_V=1.5-2.4$ mag. The southern component, which we have characterized here, appears to be the more heavily reddened ($A_V=8-12$) component of the system, and perhaps even the slightly hotter ($\sim$ 4000-6500 K) and therefore more massive ($>0.5 M_\odot$) component, depending on the adopted temperature.  Both components have infrared excess though there are no spatially resolved flux measurements beyond those at 10.7 $\mu$m that are newly reported here.

Notable is the much higher extinction required in order to explain the optical to near-infrared spectral energy distribution of FU Ori S, compared to the more moderate reddening inferred from other studies towards FU Ori N. This difference should be put in perspective with extinction estimates obtained using X-ray observations of this system. Two separate line of sight absorptions are needed to fit the FU Ori X ray spectrum to thermal emission models. Building upon earlier work with XMM-Newton \citep{2006ApJ...643..995S}, \citet{2010ApJ...722.1654S} used Chandra ACIS-S data to show that the FU Ori spectrum is fit well with (1) a highly absorbed hot plasma component ($kT \simeq 4.5$ keV; $N_H \simeq 10^{23} \; cm^{-2}$), that is spatially coincident with the primary, and (2) a cooler, less absorbed thermal plasma ($kT \simeq 0.5$ keV; $NH \simeq 10^{22}  \; cm^{-2}$) which shows a small centroid offset towards FU Ori S. They suggest that the latter soft component is most likely due to FU Ori S, which contributes at least to $50 \%$ of the soft  X-ray counts. Table \ref{tab::FuExtinction} reports the values of the estimated optical extinction for the  FU Ori N-S pair, \cite{2007ApJ...669..483Z} and this work respectively, converts them to hydrogen column density, and compares these indirect $N_H$ estimates with the direct X-ray measurement in \citet{2010ApJ...722.1654S}. Estimated hydrogen column densities in the direction of FU Ori S obtained using the two methods are of the same order of magnitude and consistent within generous error bars. The independent estimate derived from our observations thus reinforces the notion of the presence of a cool plasma embedding FU Ori  S. Moreover it could potentially be useful to externally constrain the soft X-ray component in the global fit presented in \citet{2010ApJ...722.1654S}. Such an exercise would yield to fitting the Chandra counts with a large  hydrogen column density for the hard X-ray counts, even larger than the values presented in \citet{2010ApJ...722.1654S}. This would further reveal a discrepancy between optical extinction measurements and X-Ray absorption the direction of FU Ori N even larger than the factor $>10$ exhibited in Table \ref{tab::FuExtinction}. This excess X-ray absorption indicates a large gas-to-dust ratio that could be explained by non-uniformities in the geometry of FU Ori N' s disk \citep{2007AstL...33..755K,1996ApJ...461..933K} or the presence of winds. A full understanding of this paradigm will require further research, and we refer the reader to the detailed discussion in \citet{2010ApJ...722.1654S} for a thorough discussion of potential avenues to identify such phenomena. 

The infrared excess, established here and implied by the data in \citet{2004ApJ...608L..65R} as well as the CO plus BrG emission reported by \citet{2012AJ....143...55B} (though mysteriously not present in the \citet{2004ApJ...608L..65R} spectrum),  strongly suggests that the FU Ori S is also a young star/disk system. However, the lack of cool molecular features and the implied temperature of at least $4000$ and perhaps more than $5000 \; K$  actually render FU Ori S the more massive object, in other words, the primary of the system. The observed optical and infra-red brightness ratios can then be explained by both the excess luminosity of FU Ori N due to its outburst state since the 1936 eruption, and the high extinction in the line of sight towards FU Ori S. Dust in the hot inner disk around FU Ori N cannot account for this extinction since the pair separation of $\sim 225$ AU is larger than published estimates of the outer radius of this disk \citep{2007ApJ...669..483Z}.  Perhaps, however, the pair is embedded in the same circumbinary envelope with higher extinction in either a clumpy envelope or the foreground cloud present along the line of sight to the S component than towards the N component. Indeed, the pair could be a binary in the process of fragmentation. The source may thus be analogous to the Z CMa binary system, where the optically visible source is also a candidate FU Ori star and the embedded companion is  the more massive object, called a Herbig Ae/Be star in the literature. The Z CMa system is discussed in a separate paper from the P1640 collaboration (\citep{HinkleyInPrep}). 

 \citet{2004ApJ...608L..65R,1996MNRAS.278L..23C,2005MNRAS.361..942C} and \citet{1992ApJ...401L..31B} have proposed binary interactions as a suitable trigger for the FU Ori outbursts, causing gravitational instabilities that raise accretion rates by 3-4 orders of magnitude  over normal T Tauri quiescent accretion levels. Problems with this origin of the outbursts (see \citet{2008A&A...492..735P,2010MNRAS.402.1349F}) include the short predicted rise times, the higher order multiplicity suggested as necessary for aperiodic instabilities than the mere binarity that is observed, and finally that the gravitational instability would affect the entire disk, not just in the inner fraction of an AU that is thought to be participating in the outburst. Note that in our case the separation between FU Ori N-S is too wide for FU Ori S to be responsible for the 1935 outburst.  

\section{Conclusion}

We have presented near-infrared Integral Field Spectrograph observations  of FU Orionis: our data spatially resolves the FU Ori N and FU Ori S binary components throughout the J- and H-bands, at R$\sim$45. Our astrometric analysis unambiguously establishes common proper motion FU Ori N and S form a binary pair. Our Spectral Energy Distribution analysis suggests that the southern faintest source in this system might actually be the more massive component.

Our observations allowed us to retrieve the SED of FU Ori S, which was thus far poorly constrained. In order to carry out unbiased spectro-photometric estimates in the presence of speckles we applied the damped LOCI algorithm \cite{LaurentSpeckle}, a reduction method specifically designed for high contrast science with IFS. This is the first communication reporting the high accuracy, on a faint astronomical source, of this technique pioneered by the Project 1640 team. 

We combined the P1640 data with all other spatially resolved information (R-band, K-band spectroscopy, L-band, and N-band) available from the literature or from re-analysis of previously existing data. L and N-band point are well above any conceivable reddened stellar photosphere, and must therefore be due to circumstellar dust around FU Ori S, whose average temperature we estimated at $\sim$480 K. The spectral energy distribution of FU Ori S is very red from optical wavelengths through the H-band with  a turnover thereafter such that the K-band spectral slope is blue. This combined data set  is best explained by an underlying star in the $4000-5000$ K temperature range behind $8-12$ mag of visual extinction. 

This value of the extinction is in good agreement with estimates obtained using previously published  X-ray measurements.  In particular our estimate is consistent with the hydrogen column density derived using Chandra soft X-ray counts, which have been mostly attributed to FU Ori S \cite{2010ApJ...722.1654S}. Our independent constraint on the extinction in the direction of FU Ori S helps in turn to further confirm the anomalously large excess hard X-ray absorption towards FU Ori N. 

The high visual extinction in the line of sight of FU Ori S, most likely due to the geometry of a potential circumbinary dust shell, and is moreover responsible for the relatively high contrast ratio with respect to FU Ori N. Our SED analysis allows us to combine estimates and of visual extinction and effective temperature, and yields values of the latter twice larger than previously published.  FU Ori S is thus much more massive than previously believed and, with a lower bound for the mass estimate of $> 0.5 M_{\bigodot}$, it is probably the most massive component of this  system. This source may thus be analogous to the Z CMa binary system. A more detailed comparative analysis of these two objects, putting in perspective the quiescent and outburst near infrared SEDs of  each components in both FU Orionis and Z CMa, is a promising avenue to better unravel role of binarity in eruptive protostars. 

\section*{Acknowledgements}
Project 1640 is funded by National Science Foundation grants AST-0520822, AST-0804417, and AST-0908484. This work was partially funded through the NASA ROSES Origins of Solar Systems Grant NMO710830/102190, the NSF AST-0908497 Grant. The adaptive optics program at Palomar is supported by NSF grants AST-0619922 and AST-1007046. Some of the research research described in this publication was carried out at the Jet Propulsion Laboratory, California Institute of Technology, under a contract with the National Aeronautics and Space Administration. LP was supported by an appointment to the NASA Postdoctoral Program at the JPL, Caltech, administered by Oak Ridge Associated Universities through a contract with NASA. LP and SH performed this work in part under contract with the California Institute of Technology (Caltech) funded by NASA through the Sagan Fellowship Program. This work was based in part on data collected at Subaru Telescope and obtained from the SMOKA, which is operated by the Astronomy Data Center, National Astronomical Observatory of Japan.

\clearpage

\begin{table}
 \caption{RELATIVE ASTROMETRY OF THE FU ORIONIS SYSTEM} 
\begin{center}
\begin{tabular}{cccc} 
 \hline
 \hline
Epoch  &  $\Delta RA $ & $\Delta DEC$ &  Reference  \\
 \hline
Oct 27 th 2002 & $0.16'' \; \pm 0.03''$ & $-0.47'' \; \pm 0.03'' $& \citet{2004ApJ...601L..83W} \\ 
 \hline
Dec 15 th 2003 & $0.147'' \; \pm 0.006''$ & $-0.470'' \; \pm 0.006''$& \citet{2004ApJ...608L..65R} \\ 
 \hline
Feb 20 th 2005 & $0.139'' \; \pm 0.017''$ &$-0.454'' \; \pm 0.017''$ & \citet{2009ApJ...700..491M} \\ 
 \hline
Jan 23 rd 2008 & $0.137'' \; \pm 0.008''$ & $-0.474'' \; \pm 0.008''$ & \citet{2008AstBu..63..357K} \\
 \hline
Mar 17 th 2009 & $0.158'' \; \pm 0.009''$&$-0.465'' \; \pm 0.009''$&This work \\ 
 \hline
Dec 12 th 2011 & $0.15'' \; \pm 0.02''$&$-0.45'' \; \pm 0.02''$& This work \\  
\end{tabular}
\label{tab::FuAstrometry}
\end{center}
\end{table}

\begin{table}
 \caption{PHOTOMETRY OF THE FU ORIONIS SYSTEM} 
\begin{center}
\begin{tabular}{ccccc} 
 \hline
 \hline
Filter  &  $\lambda_c$ &FU Ori N &   FU Ori S & Instrument  \\
 \hline
J & $1.250$ & $6.519 \pm 0.015$ & $11.98 \pm 0.19$ & P1640$^{(a)}$\\
 \hline
J & $1.250$ & $6.519 \pm 0.015$ & $11.69 \pm 0.22$ & IRCS $^{(b)}$ \\
 \hline
H & $1.635$ &  $5.699 \pm 0.029$ & $10.36 \pm 0.17$ & P1640 $ ^{(a)}$\\
 \hline
H & $1.635$ &  $5.699 \pm 0.029$ & $10.14 \pm 0.19$ & IRCS  $^{(b)}$\\
 \hline
K' & $2.121$ &  $5.259 \pm 0.023$ & $9.35 \pm 0.15$ &  IRCS  $^{(b)}$\\
 \hline
L' & $3.770$ &  $4.180  \pm 0.039$ & $8.09 \pm 0.16$ &  IRCS  $^{(b)}$\\
  \hline
N & $10.7$ &  $  2.2  \pm 0.02 $ & $ 4.7 \pm 0.13$ &  Keck Segment masking  $^{(c)}$\\
 \hline
\end{tabular}
 \label{tab::FuPhotometry}
  \note{(a) Band averaged photometry based on the P1640 spectra.\\(b) IRCS program in \citet{2004ApJ...608L..65R} reprocessed in this work.\\(c) Data from \citet{2009ApJ...700..491M} reprocessed in this work.}
\end{center}
\end{table}

\begin{table}
 \caption{EXTINCTION IN THE FU ORIONIS BINARY PAIR} 
\begin{center}
\begin{tabular}{cccc} 
 \hline
 \hline
Line of sight & $A_V$ in mag $^{(a)}$ & $N_H =f(A_V)$ in $10^{22}$ cm$^{-2}$ $^{(b)}$ &  $N_H$ in $10^{22}$  cm $^{-2}$ $^{(c)}$   \\
 \hline
FU Ori S & $ 10 \pm 2$  & $2 \pm 0.4 $& $1.1 \pm 0.5 $\\
 \hline
FU Ori N & $2 \pm 0.5$   & $0.39 \pm 0.11$  & $10 \pm 5$\\
\hline
\end{tabular}
\label{tab::FuExtinction}
\note{(a) \citet{2010ApJ...713.1134Z} for FU Ori N, this work for FU Ori S. \\(b) Conversion carried out using $ftp://ftp.astro.princeton.edu/draine/dust/mix/kext_albedo_WD_MW_3.1_60$.\\(c) Best fit for the two components model in \citet{2010ApJ...722.1654S}.}
\end{center}
\end{table}

\end{document}